\documentclass[twocolumn,preprintnumbers,amsmath,amssymb]{revtex4-2}
\usepackage{graphicx}
\usepackage[normalem]{ulem}
\usepackage[svgnames]{xcolor}
\usepackage{bm}
\usepackage{multirow}
\usepackage{float}
\usepackage{titlesec}
\usepackage[columnwise]{lineno}
\usepackage{float}
\usepackage{titlesec}
\usepackage{gensymb}
\usepackage{textcomp}

\begin{document}



\title{The polarisation fluctuation length scale shaping the superconducting dome of SrTiO$_3$}
\author{Beno\^{\i}t Fauqu\'e$^{1}$}
\author{Shan Jiang$^{1}$}
\author{Tom Fennell$^{2}$}
\author{Bertrand Roessli$^{2}$}
\author{Alexandre Ivanov $^{3}$}
\author{Celine Roux-Byl$^{4}$}
\author{Benoît Baptiste$^{5}$}
\author{Philippe Bourges$^{6}$}
\author{Kamran Behnia$^{4}$}
\author{Yasuhide Tomioka$^{7}$}

\affiliation{
$^{1}$ JEIP (USR 3573 CNRS), Coll\`ege de France,  75005 Paris, France\\
$^{2}$ Laboratory for Neutron Scattering and Imaging, Paul Scherrer Institut, Villigen, Switzerland \\
$^{3}$ Intitut Laue Langevin, 71 avenue des Martyrs, CS 20156, 38042 Grenoble Cedex, France\\
$^{4}$ Laboratoire de Physique et d'Etude de Mat\'{e}riaux (CNRS)\\ ESPCI Paris, Université PSL, 75005 Paris, France\\
$^{5}$ IMPMC-Sorbonne Université and CNRS, 4, place Jussieu, 75005 Paris, France\\
$^{6}$ Laboratoire L\'eon Brillouin, CEA-CNRS, Universit\'e Paris-Saclay, CEA Saclay, 91191 Gif-sur-Yvette, France\\
$^{7}$ National Institute of Advanced Industrial Science and Technology (AIST), Tsukuba, 305-8565, Japan\\
}

\date{\today}
\begin{abstract}
Superconducting domes, ubiquitous across a variety of quantum materials, are often understood as a window favorite for pairing opened by the fluctuations of competing orders. Yet, a quantitative understanding of how such a window closes is missing. Here, we show that inelastic neutron scattering, by quantifying a length scale associated with polar fluctuations, $\ell_0$, addresses this issue. We find that the superconducting dome of strontium titanate definitely ends when $\ell_0$ vanishes. Moreover, the product of $\ell_0$ and the Fermi wavevector peaks close to the maximum critical temperature. Thus, this superconducting dome stems from the competition between the increase of the density of states and the unavoidable collapse of the quantum paraelectric phase, both induced by doping. The successful quantitative account of both the peak and the end of the superconducting dome implies a central role in the pairing mechanism played by the soft ferro-electric mode and its hybridisation with the acoustic branch. Such a scenario may also be at work in other quantum paraelectric materials, either bulk or interfaces.
\end{abstract}

\maketitle

\section{Introduction}

In many superconducting materials such as cuprates \cite{Keimer2015}, pinictides \cite{Fernandes2022}, heavy fermions \cite{White2015}, doped band insulators \cite{bustarret2015}, and oxide \cite{Zubko2011} or graphene \cite{Balents2020} heterostructures, the superconducting temperature transition (T$_c$) displays a dome shape as a function of the doping. Identifying the mechanisms giving rise to these domes is the subject of intense debates.

Among superconducting domes, the case of doped SrTiO$_3$ is unique (see \cite{CollignonRev2019,GASTIASORO} for reviews). In bulk crystals, the superconducting dome starts at a carrier concentration as low as n $\simeq 5\times 10^{17}$cm$^{-3}$ for oxygen reduced samples \cite{Lin2013,Yoon2021}, spans over more than three orders of magnitude and ends at n $\simeq 10^{21}$cm$^{-3}$, irrespective of the dopant identity \cite{Tomioka2019,Tomioka2022}. This corresponds to a doping level of about 5$\%$, an extremely low doping level compared with the threshold of superconductivity in many other matrials. Moreover, in contrast to other families, the parent compounds is not magnetic but is a {\emph{quantum paraelectric}}. This state, characterized by a large dielectric constant $\epsilon \approx$ 2 $\times$ 10$^{4}$ \cite{Muller1979}, is driven by the softening of zone centered transverse optical (TO) phonon mode that hybridises with the transverse acoustic (TA) phonon mode at low temperature \cite{Yamada1969,Courtens1993,Delaire2020,Fauque2022}. 

Recently, it has been suggesting that the coupling of the electrons with ferro-electric fluctuations can lead to superconducting pairing, in particular down to very low density. This coupling can take either the form of quadratic coupling \cite{Marel2019,Kiselov2021,Volkov2022} or a Rashba type electron-TO mode coupling \cite{Yoon2021,Yu2022,Gastiasoro2022}. In these theories, the attractive pairing mechanism is static with an amplitude set by $k_F$, the Fermi wave-vector and $\ell_{0}$, the spatial length scale of the zero-point energy polarisation fluctuations. $\ell_{0}$ is given by the TO dispersion energy ($E$) satisfying in the parent compound \cite{Yamada1969,Courtens1993} : 
\begin{equation}
    E^2(q)=\omega^2_{TO}[1+\ell^2_{0}q^2]
    \label{EqTO}
\end{equation} where $\ell_{0}$=$\frac{v_{TO}}{\omega_{TO}}$  with
$\omega_{TO}$ and $v_{TO}$ are the energy and velocity of the TO mode at the zone center.

\begin{figure*}
\centering
\makebox{\includegraphics[width=1.0\textwidth]{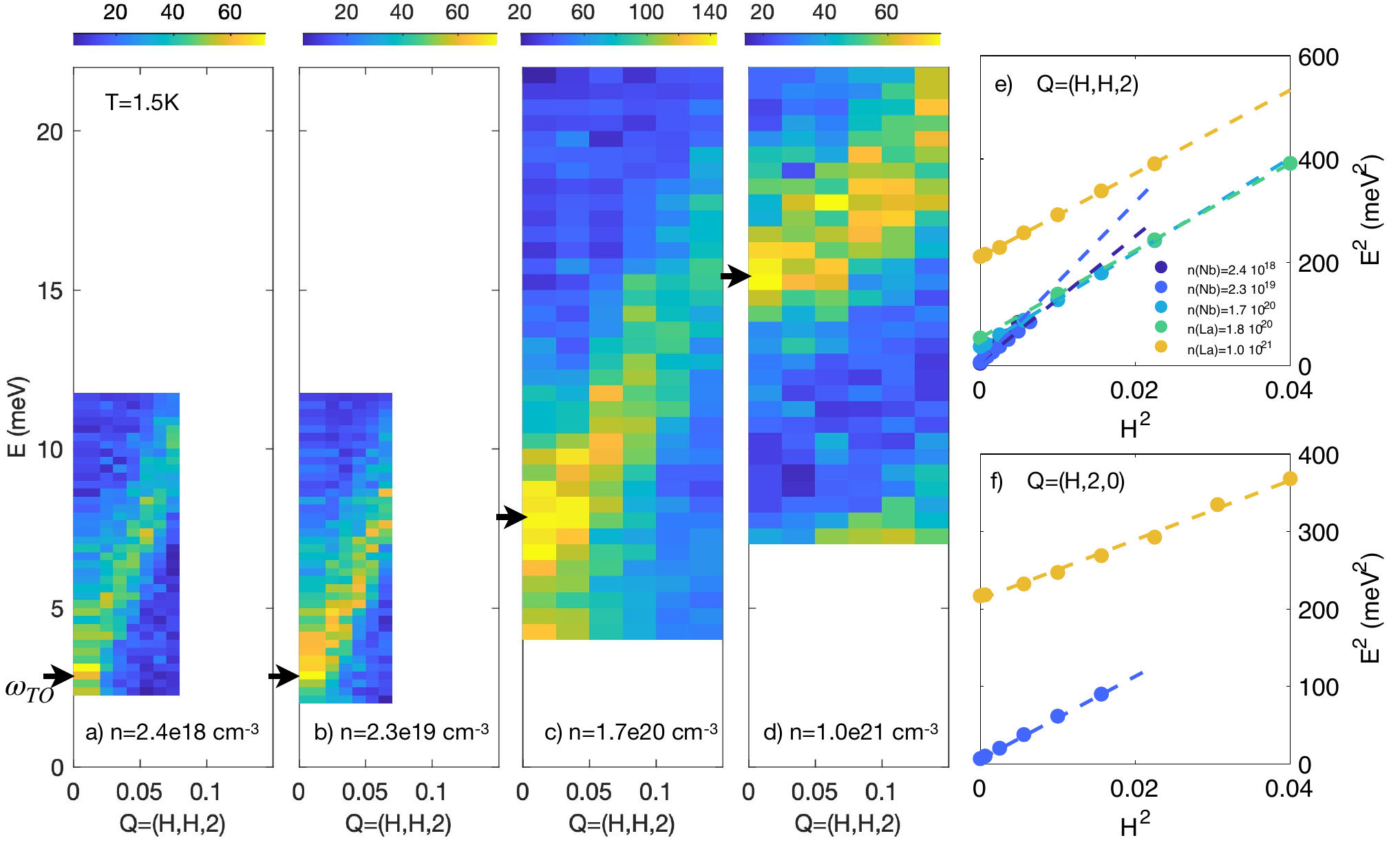}}
\caption{{\textbf{Doping evolution of the soft ferroelectric mode dispersion at T = 1.5 K:}} Momentum-and energy-resolved neutron scattering intensity map of along the (H,H,2) direction in SrTi$_{1-x}$Nb$_x$O$_3$ with an electron density a) n(Nb)=2.4$\times$10$^{18}$ cm$^{-3}$, b) n(Nb)=2.3$\times$10$^{19}$cm$^{-3}$ in Sr$_{1-x}$La$_x$TiO$_{3}$ with c) n(Nb)=1.8$\times$10$^{20}$cm$^{-3}$ and  d) n(La)=1.0$\times$10$^{21}$cm$^{-3}$ at T = 1.5 K. e) Dispersion of the soft ferro-electric mode along [H,H,0] for the five doping studied as function of H$^2$. f) same as e) along [H,0,0]. For all the doping the dispersion can be fitted through : $E^2(q)$=$\omega^2_{TO}$+$v^2_{TO}q^2$, where q[1,1,0]=$\frac{2\pi}{a}\sqrt{2}$H and q[1,0,0]=$\frac{2\pi}{a}$H where $a$ is the lattice parameter of SrTiO$_3$, that allows one to determine $\ell_0=\frac{v_{TO}}{\omega_{TO}}$. Black arrows on a-d) indicate the position of $\omega_{TO}$.}
\label{FigRaw}
\end{figure*}

\begin{figure*}
\centering
\makebox{\includegraphics[width=1.0\textwidth]{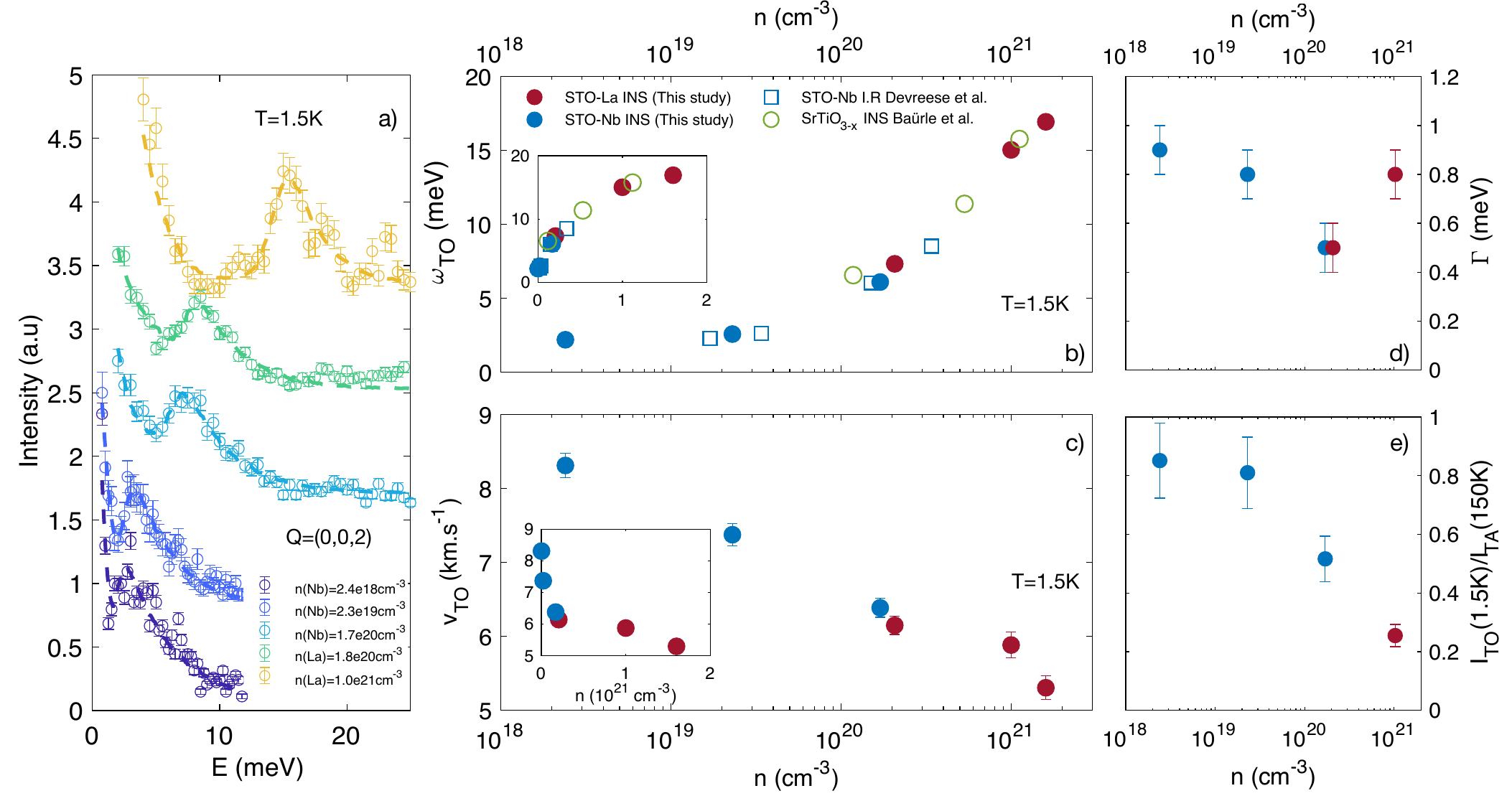}}
\caption{{\textbf{Doping evolution of the soft ferro-electric mode properties at T = 1.5 K: } a) E-scans at ${\bf{Q}}=(0,0,2)$ for five electron doped samples. Curves are shifted for clarity. b) $\omega_{TO}$ {\it{vs.}} $n$ in linear-log scale in oxygen reduced STO (green points from \cite{Bauerle1980}), in Nb-doped STO (close circle points are INS data, open squares are from infrared measurements from \cite{Devreese2010}), in La-doped STO (close red circles). Inset :  $\omega_{TO}$ {\it{vs.}} $n$ in linear scale. c) $v_{TO}$ {\it{vs.}} $n$ in linear-log scale in Nb and La doped STO deduced from the fit of the dispersion shown in Fig. \ref{FigRaw}. Inset : $v_{TO}$ {\it{vs.}} $n$ in linear scale d) Energy width ($\Gamma$) of the TO mode at T = 1.5 K. e) Intensity of the TO mode normalised with the TA mode intensity at T = 150 K and ${\bf{Q}}$=(0.05,0.05,2). Error bars are deduced from the convolution fitter that finds the best least-squares fit of the dispersion parameters (see Supplementary Material section B). }}
\label{FigDoping}
\end{figure*}

\begin{figure}[ht!]
\begin{center}
\includegraphics[angle=0,width=8cm]{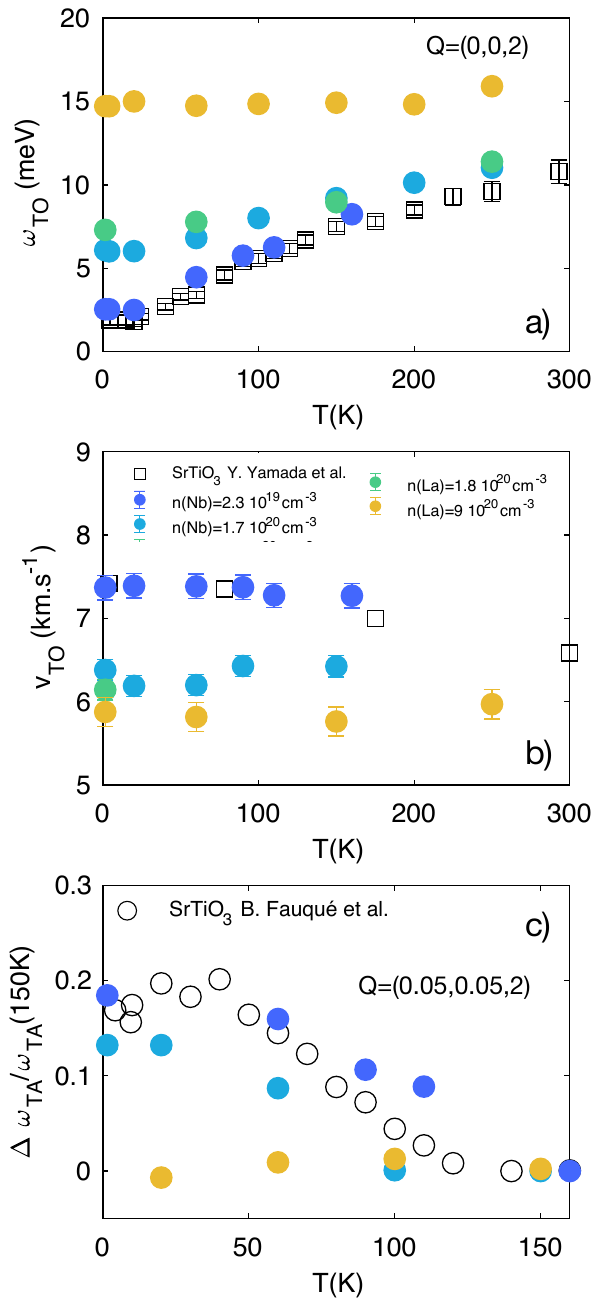}
\caption{\textbf{Temperature dependence of the soft ferroelectric mode:} a) and b) Temperature dependence of $\omega_{TO}$ and $v_{TO}$ for the five doping studied compare with the parent compound values from \cite{Yamada1969} (in black open circle points). c) Amplitude of the TA softening at $\bf{Q}$=(0.05,0.05,2) versus temperature for three dopings compare with the parent compound behavior (in black open square points from \cite{Fauque2022}). As the TO mode harder the TA softening vanishes consequence of the decrease of the coupling between both modes.}
\label{FigTemp}
\end{center}
\end{figure}

\begin{figure}[ht!]
\begin{center}
\includegraphics[angle=0,width=9cm]{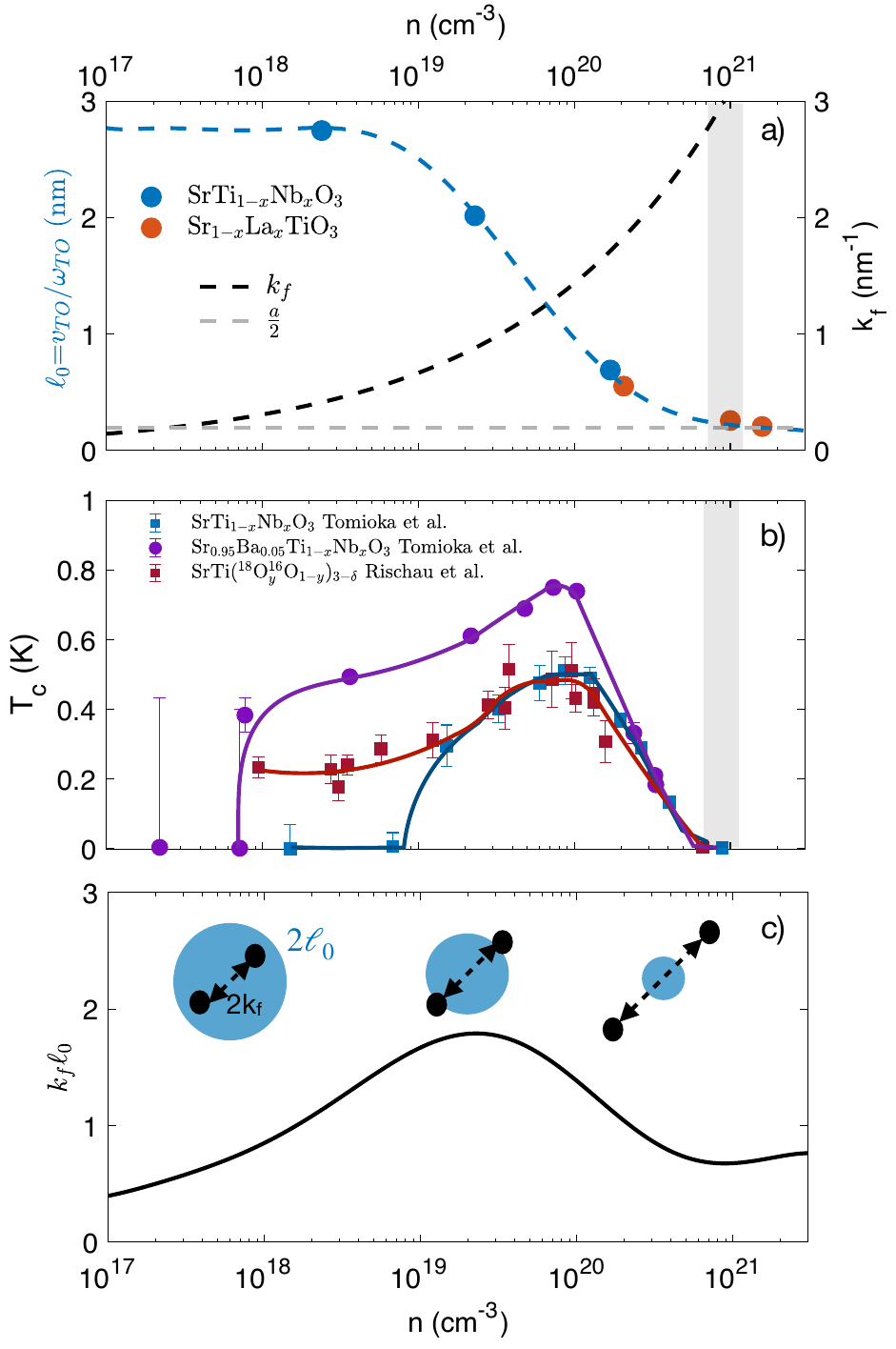}
\caption{{\textbf{$\ell_0$, T$_c$ and $k_F$:}} a) Doping evolution of $\ell_0$=$\frac{v_{TO}}{\omega_{TO}}$ (in blue and red points), deduced from Fig.\ref{FigDoping}, $k_F=(3\pi^2n)^\frac{1}{3}$ (in black dot line) and $\frac{a}{2}$ (in gray dot line) where $a$ = 3.9 $\AA$ is the lattice parameter of SrTiO$_3$. b) Doping evolution of the superconducting critical temperature ($Tc$) according to \cite{Tomioka2019,Rischau2017,Tomioka2022}. With electron doping $\ell_0$ decreases and saturate to about $\frac{a}{2}$ while $k_F$ increasing. The gray areas in a and b) mark the end of the superconducting dome concomitant with the saturation of $\ell_0$ to about $\frac{a}{2}$. c) $k_F\ell_0$ {\it{vs.}} $n$ and displays a dome shape. Inset : sketch of the electron doping evolution of $2\ell_0$ (blue disk) and 2$k_F$ (black dot arrows). The summit of the superconducting dome occurs when $k_F\ell_0$ is the largest, see the text.}
\label{FigDome}
\end{center}
\end{figure}

Here, we present the first systematic study of the dispersion of the soft ferro-electric phonon mode in electron doped SrTi$_{1-x}$Nb$_x$O$_{3}$ and Sr$_{1-x}$La$_x$TiO$_{3}$. Using inelastic neutron scattering (INS), a probe that allows a determination of  $\ell_{0}$, we track its decrease with doping as a result of the hardening of the TO-mode, the decrease of the speed of the TO mode and the breakdown of the TO/TA hybridisation. Comparing $\ell_0$ and $k_F$ unveils the two length scales shaping the superconducting dome of SrTiO$_3$. With the electron doping, $k_F$ increases while $\ell_0$ shrinks and saturates to half of the lattice parameter value at n$\simeq 10^{21}$cm$^{-3}$. Thus, the end of the superconducting dome and the breakdown of the quantum paraelectric state are concomitant. Our result provides direct evidence of the key role of the soft TO mode and its hybridisation with the TA mode in the superconducting dome of quantum paraelectrics.

\section{Results}

Fig.\ref{FigRaw}a-d) show the dispersion of the soft ferro-electric mode as measured by INS along the (H,H,0) direction for carrier densities ranging from 2.4$\times$10$^{18}$cm$^{-3}$ to 1.0$\times$10$^{21}$cm$^{-3}$ at  T = 1.5 K. The identity of the dopants (La or Nb), is indicated in parentheses for each carrier density. Description of the samples, their characterisation by electrical transport measurements and the spectrometers configurations are given in the Supplementary Material section A. Fig.\ref{FigDoping}a) shows the energy scans at $\bf{Q}$=(0,0,2) for five of the samples studied at T = 1.5 K. The soft TO mode begins to harden when the carrier density exceed 2.3$\times$10$^{19}$cm$^{-3}$. 

Convolution of the phonon spectral function with the resolution spectrometer function (see Supplementary Material section B), allows accurate fits and independent determination of $\omega_{TO}$, $v_{TO}$ and the energy width of the TO-mode ($\Gamma$). In the whole doping range, the energy dispersion of the TO mode satisfies Eq.\ref{EqTO} along both directions (H,H,2) and (H,2,0), see Fig.\ref{FigRaw}e)-f). 

Fig.\ref{FigDoping}b-d) show the deduced doping evolution of $\omega_{TO}$, $v_{TO}$ and $\Gamma$. As function of the doping $\omega_{TO}$ increases from 2.1 meV at $n$(Nb)=2.4$\times$10$^{19}$cm$^{-3}$, comparable with the value in pure SrTiO$_3$, to 17.0 meV at $n$(La)=1.6$\times$10$^{21}$cm$^{-3}$. This sudden hardening of the TO mode is accompanied by a slight decrease of $v_{TO}$ at low temperature of about 40$\%$ 
and a slight decrease of its energy width. We note, like in pure SrTiO$_3$ \cite{Yamada1969,Courtens1993}, no anisotropy in the TO-dispersion across the doping range studied ($v_{TO}$(${\bf{Q}}//[1,1,0]$)$\approx$ $v_{TO}$(${\bf{Q}}//[1,0,0]$).

The collapse of the quantum para-electric regime upon doping is further supported by the temperature dependence of $\omega_{TO}$ and $v_{TO}$ shown on Fig.\ref{FigTemp}a and b). Up to $n$(Nb)=2.4$\times$10$^{19}$cm$^{-3}$ the T-dependence of $\omega_{TO}$ and $v_{TO}$ show almost no difference from the parent compound \cite{Yamada1969} (black open square points in Fig.\ref{FigRaw} a) and b)). Above this concentration, the TO mode becomes almost temperature independent. At the highest doping studied, the system is no more in the quantum para-electric phase. $v_{TO}$, in contrast, barely changes with temperature and decreases by 40$\%$ with doping. 


 Previous studies documented hybridization between the TO and TA branches in the undoped SrTiO$_3$ \cite{Yamada1969,Courtens1993,Fauque2022}. This hybridisation, a consequence of the anharmonic coupling phonon modes \cite{Delaire2020}, is concomitant with the softening of the TO mode. It manifests itself in two ways. First, there is a dramatic transfer of intensity from the TA branch toward the TO mode ${\bf{Q}}$=(0,0,2) \cite{Yamada1969,Delaire2020}. Second, the TA  mode itself softens at a finite wave-vector \cite{Fauque2022}. The latter feature has been recently ascribed to a flexo-electric effect \cite{Verri2023}. According to our results, doping not only hardens the  TO mode, but also weakens the TO-TA hybridization. With increasing  doping, the absolute intensity of the TO mode, normalised by the amplitude of the TA mode at high temperature in each samples (see Supplementary Material section B), decreases by about a factor four (see Fig.\ref{FigDoping}e)). Simultaneously, the TA softening, measured at ${\bf{Q}}$=(0.05,0.05,2), decreases and vanishes at the highest doping (see Fig.\ref{FigRaw}c)). Thus, the hardening of the TO mode is accompanied with the weakening of its coupling with the TA mode. As a result, the spectral weight transfer and  the softening of the TA mode both eventually vanish with doping. The breakdown of the quantum para-electric regime is thus concomitant with the breakdown of the TO-TA hybridization, providing another evidence that both effects are intimately related.

 \section{Discussion}


Our measurement allows us to quantify, for the first time, the doping evolution of $\ell_0=\frac{v_{TO}}{\omega_{TO}}$, (see Fig.\ref{FigDome}a). Due to the combined change of $\omega_{TO}$ and $v_{TO}$, $\ell_0$ decreases by more than one order of magnitude with the doping and saturates to $\frac{a}{2}$, where $a$ is the lattice parameter, above $n$ $\approx$ 1$\times$10$^{21}$ cm$^{-3}$. The comparison of the doping evolution of $\ell_0$ with T$_c$ is a clue to the end of the superconducting dome of SrTiO$_3$. Fig.\ref{FigDome}b) shows the superconducting domes for SrTi$_{1-x}$Nb$_x$O$_{3}$, Sr$_{1-x}$La$_x$TiO$_{3}$ \cite{Tomioka2022} and SrTi($^{18}$O$_{y}^{16}$O$_{1-y}$)$_{3-\delta}$ \cite{Rischau2017}. No matter the nature of the dopants, all the domes end at a carrier density of $n \approx$ 1$\times$10$^{21}$ cm$^{-3}$. The end of the superconducting domes is thus concomitant with the saturation of $\ell_0 \approx \frac{a}{2}$, i.e. the end of the quantum-para-electric regime. We note that our doping evolution of $\omega_{TO}$ is in good agreement with early infra-red measurements \cite{Devreese2010} in SrTi$_{1-x}$Nb$_{x}$O$_3$ and INS measurements in oxygen reduced SrTiO$_{3}$ \cite{Bauerle1980}. 


The key role of $\ell_0$ in shaping the superconducting dome of SrTiO$_3$ is further highlighted once it is compared with the density of states length scale, $k_F$. For Nb and La doped samples the evolution of the Fermi surface is well captured by the rigid band approximation \cite{Allen2013,Fauque2023}. It is formed by three non-parabolic bands located at the $\Gamma$ point that are successfully fill. For sake of simplicity, we will consider the case of a single isotropic band where $n$=$\frac{k^3_F}{3\pi^2}$. $k_F$ {\it{vs.}} $n$ is shown on Fig.\ref{FigDome}a) with black dot line. With the carrier density increasing, $k_F$ increases, while $\ell_0$ decreases. Interestingly, $k_F\ell_0$ displays a dome shape which peaks at $n$ = 3$\times$10$^{19}$ cm$^{-3}$, the middle of the superconducting domes of doped SrTiO$_3$, see Fig.\ref{FigDome}c). This comparison allows us to draw a simple picture to the doping evolution of $T_c$, sketched in the insert of Fig.\ref{FigDome}c). At low doping, where $\ell_0$ $\approx$  3 nm, $T_c$ increases due to the increase of $k_F$ up to $n$ = 3$\times$10$^{19}$ cm$^{-3}$ where $\ell_0$ start to decrease. Above this level $\ell_0$ becomes too short to give rise to the maximum superconducting pairing amplitude and T$_c$ drops following $\ell_0$.

So far, the overdoped regime has been understood as the end of the anti-adiabatic regime \cite{yoon2021lowdensity} and as the passage from the clean to dirty limit \cite{Collignon2017}. Our result identifies the length scale $\ell_0$ which drives the decrease of $T_c$ and of the superconducting gap, $\Delta$, whose ratio has been found constant to the weak-coupling BCS value across the dome \cite{Scheffler2018,yoon2021lowdensity}. Qualitatively, this result is consistent with quantum critical theories applied to ferro-electrics \cite{Edge2015,Rowley2014,Enderlein2020} and with theoretical works where electrons are coupled with the soft TO-mode \cite{Marel2019,Kiselov2021,Volkov2022,Yu2022,Gastiasoro2022}. Combined with the known Fermi surface probed by quantum oscillations studies \cite{Uwe_1985,Allen2013,Lin2013,Fauque2023}, our result provides the parameters to quantitatively test theories of superconductivity in SrTiO$_3$ as well as it helps to understand the electron doping evolution of the unusual thermal conductivity \cite{Martelli2018}, isotropic magnto-resistance \cite{Collignon2020} and $T^2$-therm resistivity \cite{Lin2015sc,Kumar2021} in doped SrTiO$_3$.

In conclusion, we show that the end of the superconducting dome of bulk SrTiO$_3$ is concomitant with the collapse of the quantum para-electric regime. This result demonstrates the key role of the soft TO mode and its hybridisation with the TA branch in the superconductivity of quantum paraelectrics and their interface.

 \section{Acknowledgements}
 We thank M. Feigelman, R. Fernandes, D. Kiseliov, M. Gastiasoro, G. Guzm\'an-Verri, R. Lobo, P. Littlewood, A. Subedi and D. Van Der Marel for useful discussions. This work was supported by the Agence Nationale de la Recherche (ANR-18-CE92-0020-01), by Jeunes Equipes de l$'$Institut de Physique du Coll\`ege de France and by a grant attributed by the Ile de France regional council. This work was also supported by Japan Society for the Promotion of Science (JSPS)
KAKENHI Grant Number 23H01135.


%

\end{document}